\documentclass[%
 aip,
 amsmath,amssymb,
 reprint,%
]{revtex4-1}

\usepackage{graphicx}
\usepackage{dcolumn}
\usepackage{bm}
\usepackage{xcolor}

\usepackage[utf8]{inputenc}
\usepackage[T1]{fontenc}
\usepackage{mathptmx}
\usepackage{etoolbox}

\makeatletter
\def\@email#1#2{%
 \endgroup
 \patchcmd{\titleblock@produce}
  {\frontmatter@RRAPformat}
  {\frontmatter@RRAPformat{\produce@RRAP{*#1\href{mailto:#2}{#2}}}\frontmatter@RRAPformat}
  {}{}
}%
\makeatother

\DeclareMathAlphabet{\mathcal}{OMS}{cmsy}{m}{n}

\begin{document}

\preprint{AIP/123-QED}

\title{Integrated Information Decomposition Unveils Major Structural Traits of {\em In Silico} and {\em In Vitro} Neuronal Networks}
\author{Gustavo Menesse}
\email{menessegem@correo.ugr.es}
\affiliation{Department of Electromagnetism and Physics of the Matter \& Institute Carlos I for Theoretical and Computational Physics, University of Granada, 18071 Granada, Spain}
\affiliation{Departamento de F\'{\i}sica, Facultad de Ciencias Exactas y Naturales, Universidad Nacional de Asunci\'{o}n, 111451 San Lorenzo, Paraguay}
\author{Akke Mats Houben}%
\affiliation{Departament de Física de la Matèria Condensada, Universitat de Barcelona, \& Universitat de Barcelona Institute of Complex Systems (UBICS), E-08028 Barcelona, Spain}
\author{Jordi Soriano}%
\affiliation{Departament de Física de la Matèria Condensada, Universitat de Barcelona, \& Universitat de Barcelona Institute of Complex Systems (UBICS), E-08028 Barcelona, Spain}
\author{Joaquín J. Torres}
\affiliation{Department of Electromagnetism and Physics of the Matter \& Institute Carlos I for Theoretical and Computational Physics, University of Granada, 18071 Granada, Spain}

\date{\today}

\begin{abstract}

The properties of complex networked systems arise from the interplay between the dynamics of their elements and the underlying topology. Thus, to understand their behaviour, it is crucial to convene as much information as possible about their topological organization. However, in a large systems such as neuronal networks, the reconstruction of such topology is usually carried out from the information encoded in the dynamics on the network, such as spike train time series, and by measuring the Transfer Entropy between system elements. The topological information recovered by these methods does not necessarily capture the connectivity layout, but rather the causal flow of information between elements. New theoretical frameworks, such as Integrated Information Decomposition ($\Phi$-ID), allow to explore the modes in which information can flow between parts of a system, opening a rich landscape of interactions between network topology, dynamics and information. Here, we apply $\Phi$-ID on {\em in silico} and {\em in vitro} data to decompose the usual Transfer Entropy measure into different modes of information transfer, namely synergistic, redundant or unique. We demonstrate that the unique information transfer is the most relevant measure to uncover structural topological details from network activity data, while redundant information only introduces residual information for this application. Although the retrieved network connectivity is still functional, it captures more details of the underlying structural topology by avoiding to take into account emergent high-order interactions and information redundancy between elements, which are important for the functional behavior, but mask the detection of direct simple interactions between elements constituted by the structural network topology.

\end{abstract}

\maketitle

\begin{quotation}
The structural topology describes how the elements of a networked system are connected with one another. In real systems, obtaining information about these structural details is not always possible, since usually one has access only to time-evolving data of the systems' behavior. However, by using these data, it is possible to infer the causal relationship between elements and reconstruct the so-called effective topology, although in general it is not  directly related to the underlying structural details of the network under study. By using the Integrated Information Decomposition framework in combination with {\em in silico} data, we show that it is possible to maximize the extraction of structural details from effective connectivity by selecting the appropriate mode of transfer information between network elements, to later apply the gained knowledge to extract key organizational features of experimental data in {\em in vitro} neuronal networks.
\end{quotation}

\section{Introduction}
Naturally formed living neuronal networks, from the relatively simple worm {\em C.~elegans}~\cite{GHOSH2017110} up to the brain, exhibit a remarkable capacity to process information and carry out tasks in response to environmental changes. The combination of network layout, intrinsic neuronal dynamics and noise suffices to shape a highly versatile complex system able to quickly adapt and act to satisfy functional demands. Since the circuitry of neuronal networks cannot change instantaneously to meet new demands, but rather through plastic mechanisms that occur over substantial time scales, information flow and functional responses are constrained by the circuitry itself. This constraint, or relationship between information flow and network architecture,  has been still poorly investigated in the neuroscience community despite its importance. 

In this context, approaches based on dynamical systems and statistical physics have demonstrated to be powerful for understanding complex systems behavior~\cite{Bertin2021}.  However, they have not always provided a complete picture. More often, the knowledge about the properties of a dynamical regime or behavior does not necessarily provide sufficient information about the `function' (in the biological sense) of a given dynamical phenomenon.

Thus, understanding the potential functional properties of a complex neuronal network requires to know almost all the details that define it. This includes the topological structure of the layout of connections as well as the emerging local and global dynamical traits, from simple spikes to large--scale synchronization. To this end, the {\em information dynamics}~\cite{Lizier2013} of a system attempts to capture both the topological and dynamical properties that characterize it from the information that is contained on it. 

Tools such as Local Information Dynamics (LID)~\cite{Lizier2013}, Partial Information Decomposition (PID)~\cite{Barrett2015} and, more recently, {\em Integrated Information Decomposition} ($\Phi$-ID) \cite{Mediano2021}, have provided general insights into the dynamics and information content of diverse dynamical systems, from cellular automata to networks of Kuramoto oscillators~\cite{Mediano2022}. The main goal of these frameworks is to provide details on how a complex system stores, transfers, and modifies information, as well as the relationship between the parts and the whole in this informational dynamics. $\Phi$-ID is interesting in the context of complex systems in general, because it can reveal emergent and higher-order interactions within the dynamics of a system~\cite{Luppi2022}, which cannot be accessed from more classical information measures such as Time--Delayed Mutual Information (TDMI). 

In complex systems such as living neuronal networks, $\Phi$-ID can be applied to decompose the information encoded in the spontaneous spiking activity of {\em in vitro} neuronal assemblies, in the form of neuronal cultures, to decipher how different modes of information processing are distributed over the system~\cite{Varley2023}. For instance, at one extreme, one could consider that most of the measured information arises from the state of a specific part of the system (information is {\em unique} to that part), and therefore, any failure or alteration in this part would substantially alter the system's evolution. At the other extreme, different parts could provide the same information ({\em redundant information}) and, thus, the loss of those parts would not affect future states of the system. In between these extremes one could consider {\em synergistic information}, in which the information about the future states of the system is shared jointly by all parts. 
    
Unfortunately, none of the above scenarios can be distinguished by measuring solely TDMI, entropy transfer, or other commonly used information measures. Indeed, TDMI captures the information content of past  states in future ones and vice versa, but the sole knowledge of TDMI is not sufficient for understanding the system's dynamical flow of information. $\Phi$-ID proposes to decompose mutual information as a sum of {\em atoms} of information that capture which block of information is carried `uniquely' (by a specific part of the system), `redundantly' (by more than one part), or `synergistically' (by different parts of the system when considered together, as a whole).

In the present work, we use neuronal activity data from experimental recordings in neuronal cultures to examine, as a proof of concept, the capacity of $\Phi$-ID to decompose the information contained in the recordings on different atoms for the purpose of getting an insight into the structural topology of the recorded networks.
This analysis is combined with data from simulated neuronal networks that mimic the experimental ones taking advantage of the knowledge of the `ground-truth topology' of the {\em in silico} description. Our results show that the `unique' information atom contains the most important information to reconstruct the ground-truth connectivity of an {\em in silico} neuronal network, and that the inferred connectivity can be used to extract interesting information on the topology of {\em in vitro} networks. Our analysis demonstrate the usefulness of $\Phi$-ID to understand information flow in neuronal networks in relation to their underlying circuitry.

\begin{figure*}
        \centering
\includegraphics[width=\linewidth]{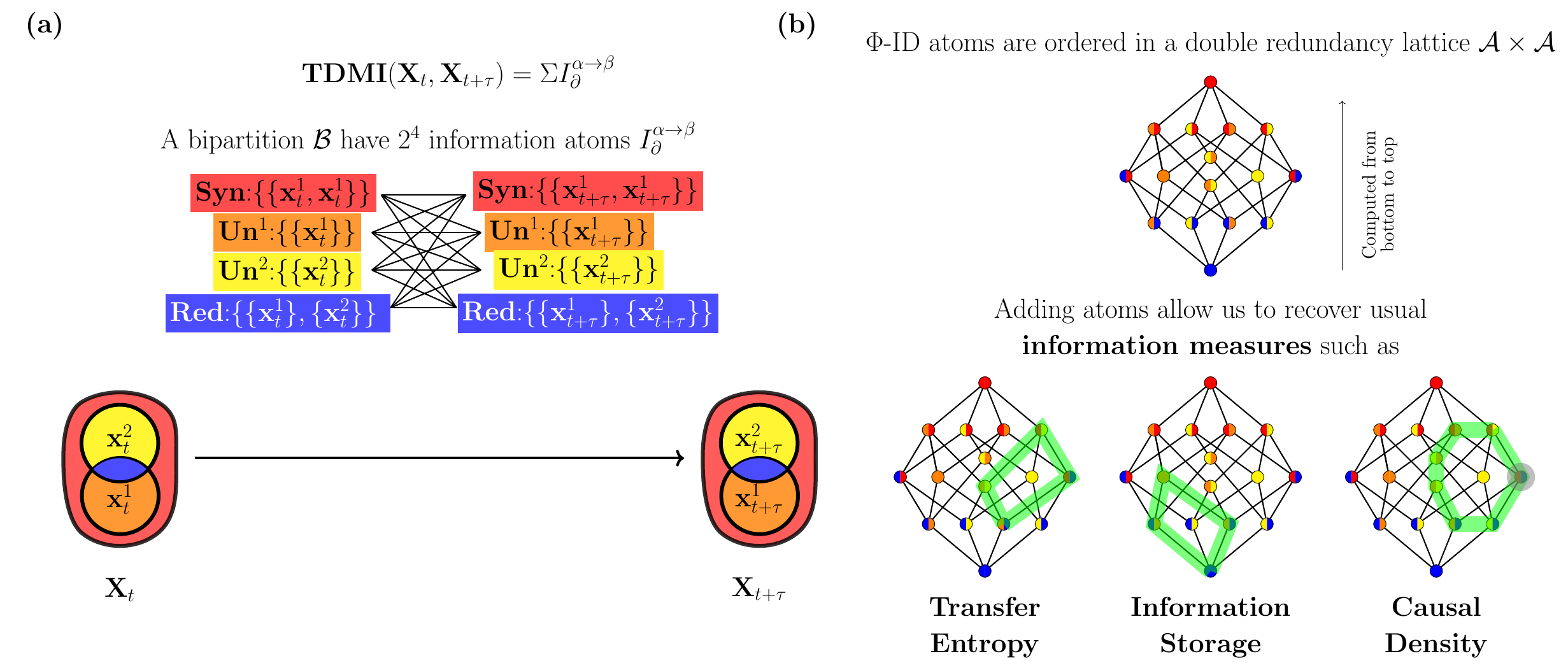}
        \caption{\textbf{Integrated Information Decomposition framework ($\Phi$-ID)}. (a) Time-delayed mutual information (TDMI) framework quantifies how the knowledge of a system's current state $\mathbf{X}_{t}$ reduces the uncertainty (i.e., provides information) about future states $\mathbf{X}_{t+\tau}$. TDMI does not provide knowledge on information flow, which is realized through $\Phi$-ID. Dividing the system $X$ in two parts such that $X=\{x^{1},x^{2}\}$, we decompose TDMI into atoms of information that capture the different modes in which information is carried by the parts of a system, namely {\em Unique} (`\textbf{Un}', orange and yellow), {\em Redundant} (`\textbf{Red}', blue), and {\em Synergistic} (`\textbf{Syn}', red). (b) Through $\Phi$-ID, each of the $2^{4}$ atoms are contained in an ordered set represented as a redundancy lattice. Each node depicts an information atom that evolves from $\alpha$ to $\beta$, represented here with the respective $\Phi$-ID atom colors. Within the $\Phi$-ID framework, other usual information measures (e.g. Transfer Entropy, Information Storage or Causal Density \cite{Seth2011}) can be decomposed into their fundamental atoms, giving the possibility to select just the relevant atoms that best suit an  application of interest.}
        \label{fig:PhiID}
    \end{figure*}

\section{Information dynamics through Integrated Information Decomposition}\label{sec:iid}
   
In the simplest case, $\Phi$-ID can be applied given a bipartition (division of system in two disconnected sub-systems) $\mathcal{B}$ of a system, such that $\mathbf{X}=\left\lbrace \mathbf{X}^{1},\mathbf{X}^{2}\right\rbrace$, and a time delay $\tau$. Mutual information is decomposed then as    
    \begin{equation}
        I(\mathbf{X}_{t},\mathbf{X}_{t+\tau}) = \sum_{\alpha,\beta\in \mathcal{A}\times\mathcal{A}}I_{\partial}^{\alpha\rightarrow \beta}  \: ,
    \end{equation}
where $\mathbf{X}_t$ represents the state of the system at time $t$. In our case, we apply $\Phi$-ID only to pairs of elements $\{x_{i},x_{j}\}$ (pairs of nodes in a neuronal network), computing the decomposition of the mutual information for each possible pair out of the $N$ elements of the network. Therefore, it should be noted that in our case, for each pair $\{x_{i},x_{j}\}$ there is only one possible bipartition.

Each information atom $I_{\partial}^{\alpha \rightarrow \beta}$ is described via a double redundancy function  $I_{\cap}^{\alpha \rightarrow \beta}$ (see below) in a double redundancy lattice $\mathcal{A} \times \mathcal{A}$, as illustrated in Fig.~\ref{fig:PhiID}. Here, $\alpha$ and $\beta$ represent the specific partial information decomposition information atom. e.g. redundant, unique or synergistic. The individual redundancy lattice is an ordered set $\mathcal{A}=\{\{ \{\mathbf{X}^{1}\}\{\mathbf{X}^{2}\}\},\{\mathbf{X}^{1}\},\{\mathbf{X}^{2}\}, \{\mathbf{X}^{1}\mathbf{X}^{2}\} \}$ as described in the Partial Information Decomposition (PID) framework\cite{Barrett2015}. Here, $\{\mathbf{X}^{1}\}\{\mathbf{X}^{2}\}$ is the \textit{redundant} (denoted as \textbf{Red}) information contained in both partition state variables. $\{\mathbf{X}^{1}\}$ and  $\{\mathbf{X}^{2}\}$ are the \textit{unique} information in each variable, symbolized as \textbf{Un$^{1}$} and \textbf{Un$^{2}$}, respectively. Finally, $\{\mathbf{X}^{1}\mathbf{X}^{2}\}$ is the information that emerges when both variables are considered together, that is, \textit{synergistic} information \textbf{Syn}. The atoms $I_{\partial}^{\alpha \rightarrow \beta}$ are defined as a difference between the redundancy and the sum of lower atoms of the double redundancy lattice,
    \begin{equation}
    I_{\partial}^{\alpha \rightarrow \beta} = I_{\cap}^{\alpha \rightarrow \beta} - \sum_{\alpha^{\prime} \rightarrow \beta^{\prime} \prec \alpha \rightarrow \beta} I_{\partial}^{\alpha^{\prime} \rightarrow \beta^{\prime}} \: .  
    \end{equation}
Here, the sum of every atom satisfies the ordering relation $\alpha^{\prime} \rightarrow \beta^{\prime} \preceq \alpha \rightarrow \beta$. The ordering relation $\preceq$ is formally defined as $\forall\, \alpha,\beta \in A, \alpha\preceq \beta \iff \forall\, b\in \beta, \exists\, a\in\alpha, a \subseteq b$. Therefore, the ordering relation of the \textit{product lattice} $\alpha^{\prime} \rightarrow \beta^{\prime} \preceq \alpha \rightarrow \beta$ means that $\alpha^{\prime} \preceq \alpha$ and $\beta^{\prime} \preceq \beta$. All atoms in the sum have the same or less redundant information as the computed atom.
  
To calculate the atoms, a double redundancy function $I_{\cap}^{\alpha \rightarrow \beta}$ must be defined. If $\alpha = \{a_1,a_2,\ldots,a_J\}$ and $\beta = \{b_1,b_2,\ldots,b_K \}$, with $\alpha,\beta \in \mathcal{A}$ and $a_j ,b_k$ non-empty subsets of $\{1,\ldots,N\}$, then this function takes the form of a set of partial information decomposition redundancies~\cite{Barrett2015} $\mathbf{Red}$, as 
    \begin{equation}
    I_{\cap}^{\alpha\rightarrow\beta} = \begin{cases}
\mathbf{Red}(\mathbf{X}_{t}^{a_1},\ldots,\mathbf{X}_{t}^{a_J};\mathbf{X}_{t+\tau}^{b_1}) & \text{if } K=1, \\
\mathbf{Red}(\mathbf{X}_{t+\tau}^{b_1},\ldots,\mathbf{X}_{t+\tau}^{b_J};\mathbf{X}_{t}^{a_1}) & \text{if } J=1, \\
I(\mathbf{X}_{t}^{a_1},\mathbf{X}_{t+\tau}^{b_1}) & \text{if } J=K=1,
\end{cases}
    \end{equation}
where $K$ and $J$ are the number of elements (parts) in sets $\alpha$ and $\beta$, respectively.

The simplest redundancy function that can be used is the minimal mutual information (MMI) \cite{Mediano2022} defined as,
\begin{equation}
    MMI = \min_{i} \,I(\mathbf{X}^{a_i}_t;\mathbf{X}^{b_1}_{t+\tau}) \: .
\end{equation}
In the present work we used MMI to implement $\Phi$-ID. More details on the mathematical construction and implementation of $\Phi$-ID are described in Mediano {\it et al.}~\cite{Mediano2021}.

An important advantage of using the $\Phi$-ID framework is the possibility of decomposing classical information measures such as Transfer Entropy (TE) into a set of core components that capture its fundamental atoms (Fig.~\ref{fig:PhiID}). In this framework, Transfer Entropy can be expressed as
\begin{equation}
    TE_{1\rightarrow 2} = I_{\partial}^{U^{1}\rightarrow R} + I_{\partial}^{U^{1}\rightarrow U^{2}}+I_{\partial}^{S\rightarrow R}+I_{\partial}^{S\rightarrow U^{2}}.
\end{equation}

\begin{figure*}
        \centering
        \includegraphics[width=.8\textwidth]{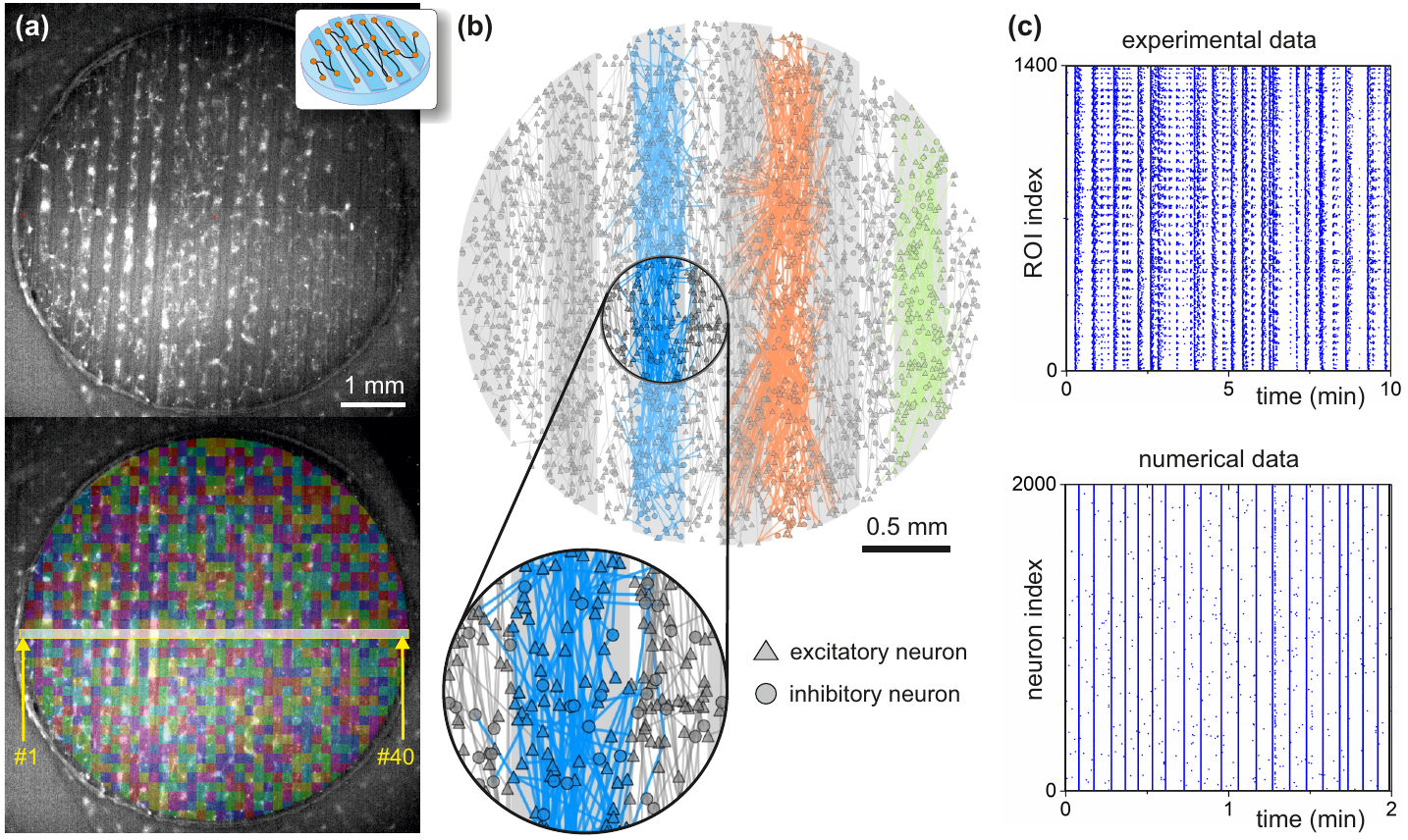}
        \caption{\textbf{Experimental and numerical data}. (a) Top: Fluorescence image of a neuronal culture grown on a topographical substrate shaped as parallel bands, as sketched in the top-right cartoon. Bright spots are active neurons. Bottom: Positioning of $\simeq 1,400$ Regions of Interest (ROIs), set as a grid of $40 \times 40$ square areas (color boxes) that cover the entire culture. 
        (b) {\em in silico} construction of the neuronal culture. 
        Neurons are placed with a higher connectivity probability along bands than across them. 
        Colors highlight neurons belonging to one vertical band and their outgoing connections. 
        The zoom in details network connectivity within a community and the presence of both excitatory (triangles) and inhibitory (circles) neurons. 
        (c) Representative raster plots of activity for experiments (top) and simulations (bottom). Each blue dot shows an activation, either within a ROI (experiments) or a neuron (simulations).}
        \label{Fig:EXP_SIMUL}
    \end{figure*}

Then, by using using MMI as redundancy function and taking $t$ and $t^{\prime}=t+\tau$ as present and future states of the variables, respectively, these atoms are written as
\begin{eqnarray}
   I_{\partial}^{U^{1}\rightarrow R} & = &  \min_{j}(I(X^{1}_t;X^{j}_{t'}))-\min_{ij}(I(X^{i}_t;X^{j}_{t'})), \label{eq:D1}\\ 
   I_{\partial}^{U^{1}\rightarrow U^{2}} & = & I({X_t}^{1};{X_{t'}}^{2}) + \min_{ij}(I(X^{i}_t;X^{j}_{t'})) \nonumber \\ & & - \min_{i}(I(X^{i}_{t};X^{2}_{t'})) - \min_{j}(I(X^{1}_t;X^{j}_{t'})), \label{eq:D2}\\
   I_{\partial}^{S\rightarrow R} & = & \min_{j}(I(\mathbf{X}_{t};X^{j}_{t'})) + \min_{ij}(I(X^{i}_t;X^{j}_{t'})) \nonumber \\ & & - \min_{j}(I(X^{1}_t;X^{j}_{t'}))  - \min_{j}(I(X^{2}_t;X^{j}_{t'})), \label{eq:D3}\\
   I_{\partial}^{S\rightarrow U^{2}} & = & I(\mathbf{X}_{t};{X_{t'}}^{2}) - I({X_t}^{1};{X_{t'}}^{2}) - I({X_t}^{2};{X_{t'}}^{2}) \nonumber \\ & & - \min_{ij}(I(X^{i}_t;X^{j}_{t'})) + \min_{i}(I(X^{i}_{t};X^{2}_{t'})) \nonumber
   \\ & & + \min_{j}(I(X^{1}_t;X^{j}_{t'})) + \min_{j}(I(X^{2}_t;X^{j}_{t'})) \nonumber \\
   & & - \min_{j}(I(\mathbf{X}_{t};X^{j}_{t'}) \:.\label{eq:D4}
\end{eqnarray}
Here, $\mathbf{X}$ means that both variables are considered jointly. Note that, by adding these four atoms, the usual definition of Transfer Entropy is recovered as conditional mutual information, i.e., 
\begin{equation}
    TE_{1\rightarrow 2} = I(X^{2}_{t^{\prime}};X^{1}_{t}|X^{2}_{t}) = I(\mathbf{X}_{t};{X_{t'}}^{2}) - I({X_t}^{2};{X_{t'}}^{2}) \:.
\end{equation}

The decomposition of Eqs.~(\ref{eq:D1})-(\ref{eq:D4}) allows to identify and isolate the relevant atoms that best suit an application of interest and to neglect those that do not significantly contribute. The quality of the selected atoms can be tested by defining a criterion for the specific application. In our case, for example, for evaluating the reconstruction of network topology from simulated data, we use the rates of true positive (TP) and false positive (FP) links to later identify which atoms provided a high TP while maintaining a low FP.

It must be noted that Transfer Entropy (TE) and its generalized version are commonly used to measure effective connectivity between nodes in a complex network\cite{stetter2012}. However, when observed from the $\Phi$-ID perspective, one realizes that just one of the atoms that compose TE is, in fact, measuring the unique information transfer between two variables. 
This atom captures the information that is unique to one variable at time $t$, and that is transferred and becomes unique to another variable at time $t^{\prime}$. 
This transfer can be understood as low order and therefore more related to a direct link or interaction between two elements $i$ and $j$ in the network, differently from synergistic information, which is an higher-order emergent information. Therefore, one may hypothesize that, in complex neuronal networks, the unique information transfer should be better for approximating the structural topology from the measured effective topology than just using TE directly. In order to demonstrate this, we have used $\Phi$-ID to infer the structure of the connections between neurons in neuronal cultures from their recorded activity.

\section{Experimental and numerical background} 

We examined the potential of using $\Phi$-ID for connectivity estimation of experimental and numerical data derived from, or inspired by, {\em in vitro} neuronal activity recordings~\cite{Soriano2023,montala2022rich} (Fig.~\ref{Fig:EXP_SIMUL}). 
As detailed in Appendix~\ref{sec:Aexpt}, experiments were prepared by growing cortical neurons, with typically 80\% excitation and 20\% inhibition, on substrates $6$~mm in diameter that incorporated a topographical modulation in the form of parallel bands ---see Fig.~\ref{Fig:EXP_SIMUL}(a)--- which effectually favored strong neuronal connectivity along bands and a weaker connectivity across them~\cite{montala2022rich}. Spontaneous activity of the culture was then recorded, and the behavior of the system was analyzed through $1,400$ Regions of Interest (ROIs) that uniformly covered the substrate. 

The numerical simulations were designed to mimic the experimental observations, by placing neurons on a bi-dimensional Euclidean space with a density and excitation-inhibition ratio similar to the one observed experimentally (see Fig.~\ref{Fig:EXP_SIMUL}(b)). 
As detailed in Appendix~\ref{sec:Anum}, connectivity between neurons was then modeled according to rules that took into account the bands-like constraints of the substrate. 
The key advantage of the synthetic networks produced in this way is that the ground-truth connectivity is known. 
This allowed to compare the inferred connections with the known structural ones, so that the usefulness of each information atom for connectivity inference could be assessed. 
Moreover, it allowed to extract interesting features of the structural connectivity, such as community structure, that naturally emerged from the underlying anisotropy (see in Fig.~\ref{Fig:EXP_SIMUL}(b) the colored groups of  neurons), and allowed comparison to the same features extracted using the inferred connectivity by the different atoms.
The procured networks were dynamically brought to life using the Izhikevich model, and the behavior of each individual neuron monitored. 

Both experiments and simulations resulted in neuronal activity data that was qualitatively similar ---see Fig.~\ref{Fig:EXP_SIMUL}(c) ---characterized by raster plots of spontaneous activity with a prominent {\em bursting} behavior, in which most of the ROIs or neurons activated together in a short time window, although they coexisted with smaller coactivations or noise-like sporadic events. These raster plots are the core data from which all information-theoretic analyses were extracted and effective connectivity quantified. Here, `effective connectivity' specifically refers to `causal influence' between pairs of nodes, in the sense that the knowledge about the present state of a node reduces uncertainty (or provides information) about the future states of the node it is interacting with~\cite{Pearl2018}.

\section{Inferring neuronal network structure using Integrated Information Decomposition}

In order to use $\Phi$-ID to infer connectivity structure from neuronal activity data, 
we calculated time-delayed mutual information (TDMI) partial atoms following the $\Phi$-ID method \cite{Mediano2021} for each pair of neurons (simulations) or ROIs (experiments). 
Then, by grouping the atoms as described in Sec.~\ref{sec:iid} we obtain the TE between each pair. Following Montal\`{a}-Flaquer \emph{et al.}~\cite{montala2022rich} and Ludl {\em et al.}~\cite{Ludl2020}, the raw measure of information of each atom and their combinations was normalized and thresholded for significance by following the z-score normalization 
\begin{equation}
    z\,_{IJ} = \frac{\text{TE}_{I\rightarrow J}-\left\langle TE_{\text{joint}}\right\rangle}{\sigma_{TE_{\text{joint}}}},
\end{equation}
where $\text{TE}_{I\rightarrow J}$ is the raw TE score between any pair of nodes $I$ and $J$, $\left\langle TE_{\text{joint}}\right\rangle$ the average of the joint distribution of all input $X$ to $J$ and output $I$ to $Y$ (for any $X$ and $Y$), and $\sigma_{TE_{\text{joint}}}$ its standard deviation. The significance for information transfer, and therefore the existence of an effective connection $I\rightarrow J$, was established by setting a threshold $z_{\text{th}}$, so that $\forall_{I,J}\; [\, z_{IJ} < z_{\text{th}} \implies z_{IJ} = 0 ]$. The remaining $z_{IJ}$ entries (significant connections) were set to $1$. The same process was performed for each partial atom and for the sum of atoms related to information transfer $I_{\partial}^{S\rightarrow U_2}+I_{\partial}^{U_1\rightarrow U_2}$. The final set of significant effective connections was stored in a matrix $A =\{ a_{IJ}\}$ for further analysis.

Such a threshold concept has been used in different studies to compare the changes in neuronal network organization in a variety of experimental and computational designs~\cite{Ludl2020,Carola2021,montala2022rich}, with the aim to describe key differences in overall network functional organization. However, as proposed by Stetter {\em et al.,}~\cite{stetter2012}, one could also consider the TE-inference approach as a way to predict the {\em structural} layout of a living neuronal culture (or its most relevant characteristics), which is in principle not completely possible. In such an effort for {\em inferring} as best as possible structural connectivity, it is clear that the choice of $z_{\text{th}}$ substantially affects the quality of reconstruction and the statistical properties of the inferred network.

Thus, considering the above arguments, and in an attempt to predict interesting structural traits of neuronal cultures, we first considered the simulated data, since its ground-truth connectivity is known, and explored all possible thresholds for each information atom, computing their corresponding Receiver Operating Characteristic (ROC) curves to evaluate the accuracy of their reconstructions. 
This allowed us to identify a range of adequate thresholds and to identify the most important atoms of information for the reconstruction of the core structural topology of the experimental data.

For the numerical simulations, and after identifying the best atom, we applied the reconstruction with different thresholds in the range that we identified as adequate.   Then, we measured important characteristics such as community structure to ascertain whether it approached the structural, ground-truth one. 
For identifying the communities we used the Louvain algorithm run in Python ({\tt NetworkX} package)~\cite{hagberg2008exploring}. In addition, for each explored threshold, we measured the distribution of number of inferred neighboring neurons as well as the distributions of euclidean distances between connected neurons. Overall, these comparisons allowed us to evaluate the impact of threshold selection on reconstruction quality.

After analyzing the numerical data, we finally  applied the same reconstruction scheme to the experimental data, by using the unique transfer information $I_{\partial}^{U^{1}\to U^{2}}$ estimate. 
We also measured the angle between effectively connected neurons in the Euclidean space to inspect whether the reconstruction captured the expected imprinted anisotropy, i.e., the presence of topographical bands on the substrate where neurons grow that impose a privileged direction for connectivity in the network.

\section{Results}

\subsection{Determination of the relevant information atoms and threshold using {\em in silico} neuronal data}

\begin{figure}
        \centering
        \includegraphics[width=\linewidth]{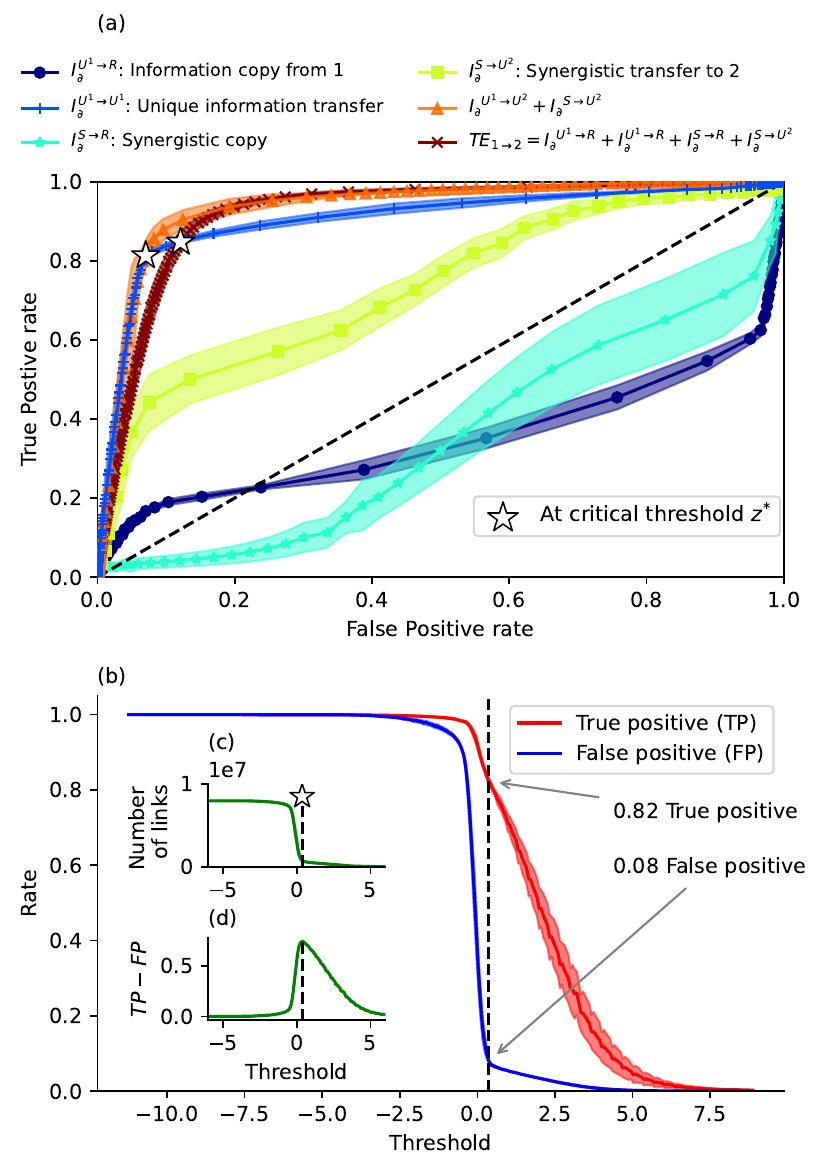}
        \caption{\textbf{Accuracy of structural topology reconstruction in {\em in silico} neuronal networks}. (a) Receiver Operating Characteristic (ROC) curves to quantify the accuracy of the reconstruction. Curves were calculated for the different information atoms that constitute the Transfer Entropy. Estimated networks are generated by including links with a calculated information score that exceeds an arbitrary threshold. ROC curves then contrast the fraction of true and false positives by comparing the inferred adjacency matrix with the ground-truth one.  ROC curves show that, for a low false positive rate, the transfer of unique information $I_{\partial}^{U_{1}\rightarrow U_{2}}$ provides the best estimate. For a higher false-positive rate, the Transfer Entropy measure has a higher true-positive rate. This effect is caused by the contribution of synergistic information transfer $I_{\partial}^{S\rightarrow U_{2}}$ (see how $I_{\partial}^{U_{1}\rightarrow U_{2}}+I_{\partial}^{S\rightarrow U_{2}}$ matches $TE_{1\rightarrow 2}$ for higher false positive values). (b) True- and false-positive rates (TP and FP, respectively) as a function of the threshold $z_{\text{th}}$ for the Unique contribution  $I_{\partial}^{U_{1}\rightarrow U_{2}}$, showing that there is an abrupt increase in FP below a critical threshold $z^* \simeq 0.4$ (vertical black dashed line). Conceptually, this abrupt change indicates that, below $z^*$, the FP rate (blue curve) quickly grows with just a small increase in TP (red curve), overall providing worse reconstructions. (c) Number of inferred links as a function of the threshold, where an abrupt change is also observed. (d) Difference between the TP and FP rate, depicting a maximum at the critical threshold $z^*$. The shadow areas indicate the standard deviation of the values in three different realization of the {\em in silico} neuronal network. The white star indicates the critical threshold point $z^*$.}. 
        \label{fig:ROC}
\end{figure}

Figure~\ref{fig:ROC}(a) shows the ROC curves for the reconstruction of the ground-truth topology of the numerical data using the different atoms that constitute the Transfer Entropy measure. We observed that the {\em unique} information transfer $I_{\partial}^{U^{1}\rightarrow U^{2}}$ ---blue line in Fig.~\ref{fig:ROC}(a)--- presents the best results for a low false positive rate (higher significance threshold $z_{\text{th}}$). This atom dominates the TE measure, exhibiting on average higher values of information than the other atoms (not shown). The other atoms introduce 'residuals' that worsen the results of the TE for low false positive rates. However, for a higher false positive rate (lower threshold), the synergistic transfer $I_{\partial}^{S\rightarrow U^{2}}$ (lime-colored line) contributes to improve the TE, which has better performance than the $I_{\partial}^{U^{1}\rightarrow U^{2}}$ alone. We demonstrate that this is the case by plotting the sum of $I_{\partial}^{U^{1}\rightarrow U^{2}} + I_{\partial}^{S\rightarrow U^{2}}$ (orange curve). We show how it matches with TE for lower thresholds (higher false positive rates), and is just slightly below $I_{\partial}^{U^{1}\rightarrow U^{2}}$ for higher thresholds (low false positive rates).

Both $I_{\partial}^{U^{1}\rightarrow U^{2}}$ and $I_{\partial}^{S\rightarrow U^{2}}$ measure the information that is only transferred to part 2, which means information that in the future $t+\tau$ becomes unique to part 2. In $I_{\partial}^{S\rightarrow U^{2}}$, the information transferred depends on higher-order interactions between parts captured by the synergy measure. In this sense, it is not just the result of a direct interaction between neurons 1 and 2, but it captures the emergent information that arises from the pair \cite{Rosas2020b}. Higher-order interactions have an impact on the causal flow of information, and therefore depart even more the effective topology from the structural one. On the other hand, the $I_{\partial}^{U^{1}\rightarrow U^{2}}$ atom captures information directly transmitted from 1 to 2, without accounting for higher-order interactions, which makes it a better approximation for capturing a direct simple interaction between both elements and, therefore, topological details more closely related to the structural aspects of the network.

Using the results for $I_{\partial}^{U^{1}\rightarrow U^{2}}$, we also plotted the values of true positive and false positive rates as a function of the threshold, as shown in Fig.~\ref{fig:ROC}(b). Here we observe that, below a `critical threshold' $z^* \simeq 0.4$, the number of false positives abruptly grows with just a small increase in true positives, substantially worsening reconstruction. Thresholds above $z^*$ smoothly reduce the amount of true positives for a very small reduction of the few remaining false positives. Indeed, just below this threshold, a large increase in the number of identified links is observed, as visible in Fig.~\ref{fig:ROC}(c). Additionally, by plotting the difference between the TP and FP rates, we observe a peak exactly at this $z^*$, as shown in Fig.~\ref{fig:ROC}(d). Thus, this threshold $z^*$ can be considered an optimal threshold in terms of the difference between TP and FP.

The point on the ROC curve of the TE reconstruction for this threshold $z^*$ is indicated by a white star in Fig.~\ref{fig:ROC}(a).
It can be seen that this point coincides with the intersection of the $I_\partial^{U^1\to U^2}$ and the TE ROC curves, meaning that for a more restrictive threshold $z>z^*$ the $I_\partial^{U^1 \to U^2}$ reconstruction outperforms the TE reconstruction.
We note that the identification of $z^*$ by plotting the number of observed links as a function of the threshold, as in Fig.~\ref{fig:ROC}(c), depends only on the recorded activity dynamics and it can be therefore also applied to \emph{in vitro}, experimental data. More specifically, the critical threshold $z^*$ appears at the right-most knee of the links-threshold curve of Fig.~\ref{fig:ROC}(c).

\begin{figure*}
        \centering
        \includegraphics[scale=0.75]{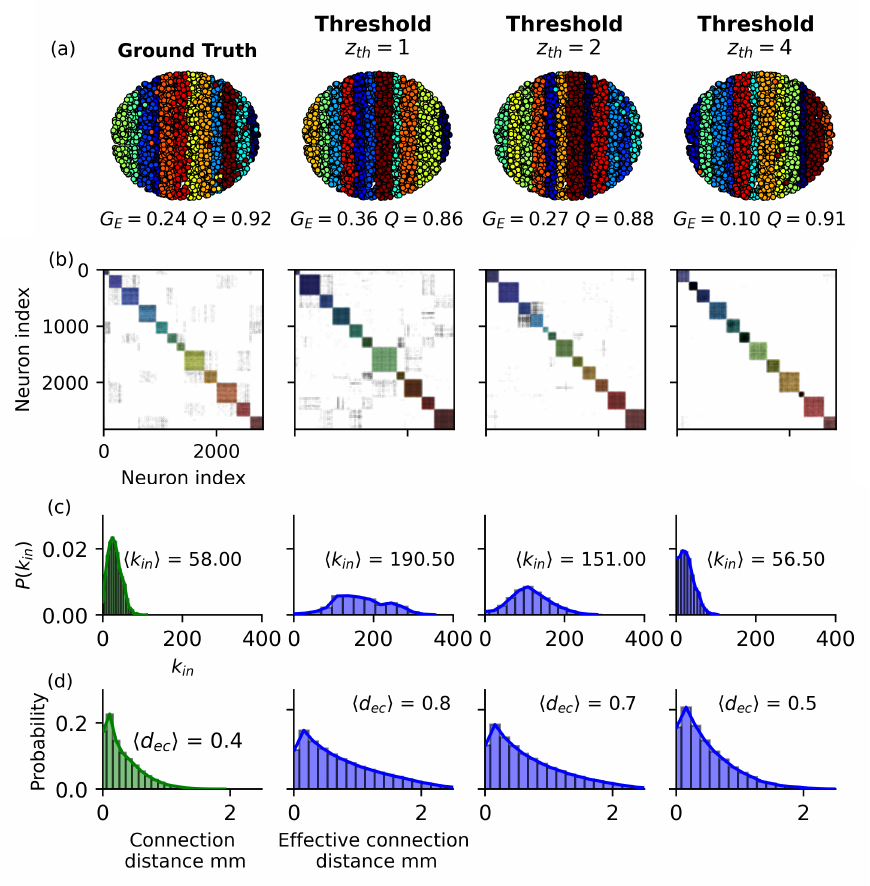}
        \caption{\textbf{Simulated network reconstruction for different thresholds}. (First column) Ground truth, structural topology used in simulations. (Following columns) Effective topology estimated using the {\em unique} atom $I_{\partial}^{U_{1}\rightarrow U_{2}}$ and thresholds $z_{\text{th}}=1,2,4$ respectively. (a) Network maps, where each dot is a neuron color coded according to the community to which it belongs. (b) Connectivity matrices. Color boxes along the diagonal of the matrices highlight the inferred communities. Neuron indices have been reordered to increase visibility. (c) Degree distribution of incoming connections $P(k_{in})$ of ground-truth (green) and reconstructed network (blue). (d) Connection distance in the ground-truth (green) and connection distance in reconstructions (blue).
        }
        \label{fig:Sim_recons}
    \end{figure*}

By focusing on the unique information transfer, we can next compare the obtained effective topology of the simulations with its ground-truth. 
Figure~\ref{fig:Sim_recons}(a) shows the ground-truth network reconstructions for different thresholds $z_{\text{th}}>z^*$. The color code indicates the community to which each node belongs. We observe that an increase in the threshold from $1$ to $4$ standard deviations does not qualitatively change the detected communities, as also revealed by a similar value of the modularity statistic $Q$. In all cases, an approximate match between ground-truth and effective network is observed. However, we regard that an increase in the threshold slightly increases the number of communities, with some of the bigger ones actually splitting into two for the highest thresholds. This gradual fragmentation of the network is expected since less connections are present overall, as captured by the global efficiency $G_E$, and therefore the reconstructed networks are expected to be more modular (higher $Q$). We also note that the threshold has an effect on spatial filtering, with higher thresholds favoring connections between closer elements, as observed previously by Montal\`{a}-Flaquer \emph{et al.}~\cite{montala2022rich} when measuring the probability distribution of information transfer grouped by Euclidean distance between elements. The detailed effect of distance (or spatial embedding of neurons) on information transfer will be explored in a future work. 

We also investigated the connectivity in the network through the degree $k$, a general property of graphs that accounts for the  number of connections of a node~\cite{medaglia2015cognitive,bassett2018nature}. Since the reconstructed effective networks are directed, one has to consider in general the incoming connections to a node (in-degree $k_{in}$) and the outgoing ones (out-degree $k_{out}$), with the total connectivity given by $k = k_{in} + k_{out}$. For clarity, Fig.~\ref{fig:Sim_recons}(c) shows only the distribution of in-degrees, i.e. the probability to observe a neuron in the network with $k_{in}$ connections. The plots show that an increase in the threshold has the expected effect of reducing the number of links per node. The ground-truth degree distribution is not recovered regardless of the threshold used, indicating that the effective topology never exactly matches the structural one. However, we observe that a higher threshold provides a better definition of a unimodal distribution, as can be seen for the thresholds $z_{\text{th}}=2$ and $4$, with the average degree matching the ground-truth for $z_{\text{th}}=4$.

Finally, Fig.~\ref{fig:Sim_recons}(d) shows the distribution of the effective connection distance, i.e. the Euclidean distance between effectively connected neurons. In general, an increase in the threshold decreases the distance, showing in fact that the threshold acts as a distance filter, favoring closer elements. Qualitatively, the distributions of the effective connection distances of the effective network agrees well with the ground-truth. However, the observed average value  in the effective network is higher than the ground-truth for all the thresholds explored. This result is expected, since the effective interaction between elements usually has a longer range than the structural connection due to the propagation of activity through intermediate nodes. Nevertheless, an additional increase of the threshold favors a closer match between distributions. However, this does not mean that the structural topology exactly agrees with the effective one in any case. Indeed, the apparent match is just an effect of filtering the longest-distance interactions in the effective topology.

\subsection{Application of $\Phi$-ID to {\em in vitro} neuronal data}

The analysis of the numerical simulations showed that the {\em unique} information transfer $I_{\partial}^{U^{1}\rightarrow U^{2}}$ was the atom that contributed the most in reconstructing the ground-truth topology. Thus, we considered this atom to extract key information of the connectivity in a living neuronal network. We considered a threshold of $z_{\text{th}}=4$ since, according to the numerical results, the statistics of the effective connectivity (average connectivity and connectivity distances) better matched the ground-truth ones. We remark that, in general, it is not possible to extract the precise topological organization of a living neuronal network from just activity data, since the neuronal connectivity layout between neurons in combination with intrinsic nonlinear neuronal dynamics and noise can  effectually lead to an infinite repertoire of dynamical states. In other words, the `inverse problem' of extracting accurate topological information from activity is not solvable. We therefore aimed here at characterizing the effective connectivity of experimental data and relate it to key features of the studied living networks such as modularity or imprinted anisotropies.

Experimental data consisted of spontaneous activity recordings of neurons grown on a topographically-modulated substrate akin to the above analyzed simulations. Three experimental repetitions were considered, which were identically prepared but resulted in slightly different networks as they matured. 

\begin{figure*}
        \centering
        \includegraphics[scale=0.75]{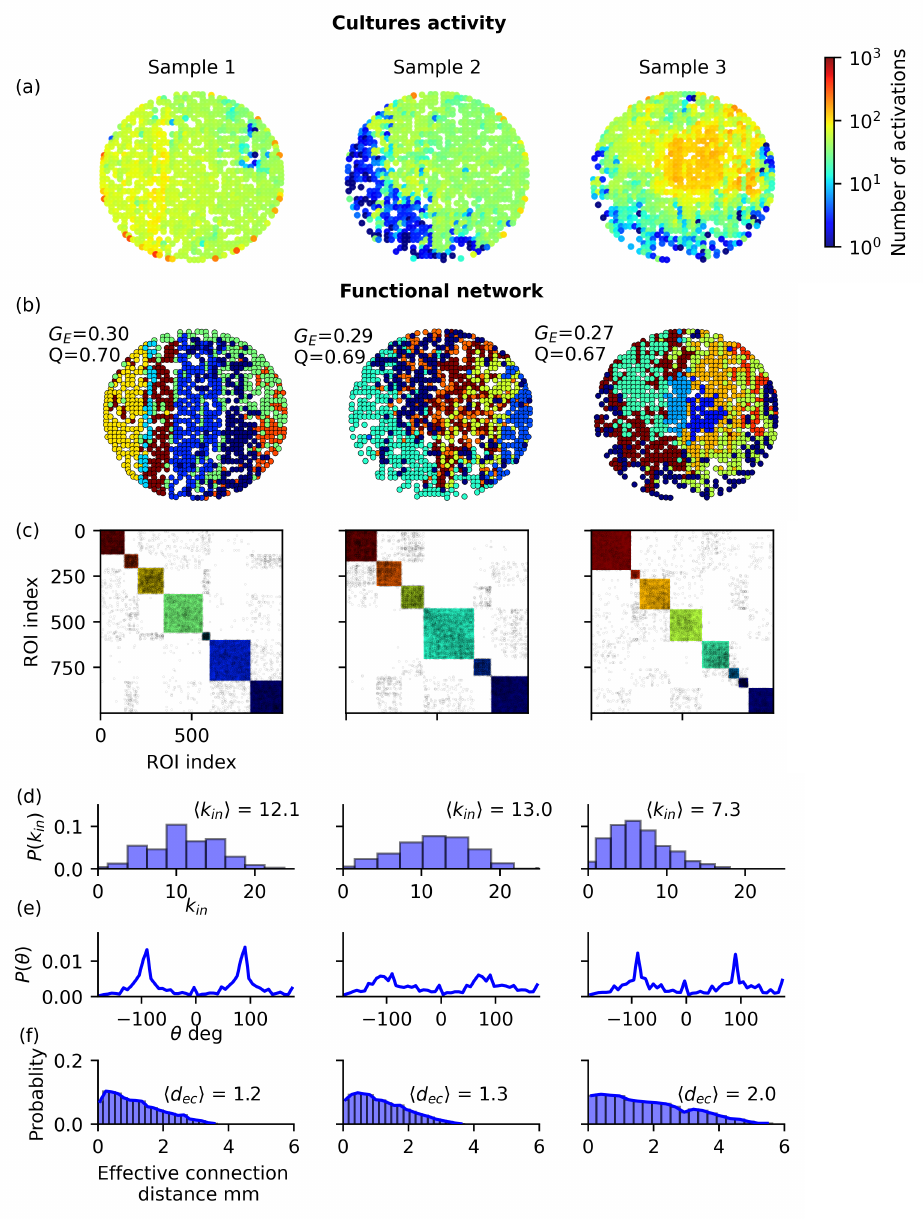}
        \caption{\textbf{Network reconstruction for three neuronal cultures $6$~mm in diameter with topographical bands}. (a) Spatial distribution of spontaneous activity in the cultures, depicting the total number of activations for each ROI in the three different cultures. (b-c) Effective connectivity maps and matrices, derived from the unique information transfer $I_{\partial}^{U^{1}\rightarrow U^{2}}$ with $z_{\text{th}}=4$. Colors identify the inferred communities. ROI indexes have been reordered to highlight community structure along the diagonal of the matrix. (d) Distribution of incoming connections $P(k_{\text{in}})$ in the reconstructed networks}. (e) Distribution of the connection angle $\theta$ in the Euclidean space. (f) Distributions of effective connection Euclidean distances $d_{ec}$ between ROIs.
        \label{fig:Exp_recons}
\end{figure*}

Figure~\ref{fig:Exp_recons} shows the results of the analysis. 
As a first observation, we note that activity was not uniform across the culture, and substantially different across repetitions ---see Figure~\ref{fig:Exp_recons}(a)---, which was caused by fluctuations in the distribution of neurons on the substrate and the intrinsic developmental variability in such a biological, living system. This lack of uniform activity had an effect on network reconstruction. Interestingly, `sample 1' had the most uniform activity across the culture and resulted in an effective connectivity that matched well the imprinted anisotropy of the topographical substrate, i.e., neuronal connectivity reflected the parallel topographical bands that shaped characteristic communities ---see Fig.~\ref{fig:Exp_recons}(b), left---. In the other samples, although communities clearly existed, they lacked such organization in bands ---see Fig.~\ref{fig:Exp_recons}(b), center and right---. Nonetheless, as shown in Fig.~\ref{fig:Exp_recons}(c), all samples had a similar number of communities and exhibited comparable bulk network properties, with a global efficiency $G_E \simeq 0.29$ and a modularity $Q \simeq 0.68$ on average. These results illustrate the difficulty in unveiling key structural traits of living neuronal networks from just activity data, and that effective connectivity cannot be arbitrarily used as a proxy of the structural blueprint of the network.

The results presented here agree with the ones presented in Montal\`{a}-Flaquer \emph{et al.} \cite{montala2022rich} where Generalized Transfer Entropy\cite{stetter2012} was applied. However, here, for the same sample (sample 1), using only the unique transfer atom, we found a larger number of stripe communities (6 here vs 4 in \cite{montala2022rich}) which is closer to the actual pattern, indicating that our reconstruction captures more details of the substrate topography. 
In addition, we found clear details of the anisotropy for all samples, while in Montal\`{a}-Flaquer \emph{et al.} \cite{montala2022rich} this property was only clearly detected in the sample with homogeneous activity (sample 1).

Figure~\ref{fig:Exp_recons}(d) compares the distributions of the number of incoming connections $P(k_{\text{in}})$ of the experimental repetitions and shows that the three samples share similar distribution shape and average $\langle k_{\text{in}}\rangle = 10.8 \pm 3.1$. This indicates that, despite the clear differences in the spatial organization of communities, the three networks share topological similarities. This is also observed in the distribution of angles between connected neurons ---see Fig.~\ref{fig:Exp_recons}(e)---, where peaks at -90$^{\circ}$ and 90$^{\circ}$ reveal that the anisotropy promoted by the bands imprints a clear directionality in connectivity, i.e., the bands funnel connections along their direction. 

To complete the analysis, we also computed the distribution of Euclidean distances between effective connections ---see Fig.~\ref{fig:Exp_recons}(f)---. We observed that samples 1 and 2 had a similar short-range connectivity, with $\langle d_{\text{ec}}\rangle \simeq 1.2$~mm, i.e. neurons effectively connected in a relatively small neighborhood that corresponded to about 20\% of the culture's diameter. Sample 3 had a markedly longer-range connectivity, with $\langle d_{\text{ec}}\rangle \simeq 2.0$~mm (a third of the culture's diameter). Such a long distance connectivity is clear in the network maps of Fig.~\ref{fig:Exp_recons}(b), where communities green and maroon horizontally extend much more than the other samples. We note, however, that these communities correspond to areas with relatively poor activity ---blue regions of Fig.~\ref{fig:Exp_recons}(a)--- indicating that they may have evolved somehow isolated from the rest of the culture. This again illustrates how small fluctuations in neuronal density or development of seemingly identical cultures have strong effects on the topological characteristics of mature networks, an aspect intrinsic to biological systems, but very difficult to model theoretically and numerically.

\section{Discussion}

Numerical simulations of living neuronal networks provide a highly valuable environment to explore the capacity of information-theoretic tools to extract major topological features ---or ideally the complete ground-truth layout--- from spontaneous activity data of the network of interest. Different studies have explored the capacity of models and frameworks to achieve such reconstruction~\cite{MAGRANSDEABRIL2018120,Banerjee2023}, and reached the conclusion that a precise reconstruction is in general unattainable given the intrinsic complex nature of a neuronal network, in which cell dynamics, topology and noise may give rise to an almost infinite repertoire of activity patterns. For instance, reconstruction is more difficult when excitatory and inhibitory neurons are considered, or when the spontaneous dynamics of the neuronal network is locked in highly correlated regimes such as network bursts~\cite{orlandi2014transfer}. Such a bursting behavior is an innate characteristic of biologically-realistic neuronal networks, both synthetic and living. Thus, it cannot be easily suppressed and has to be treated as a feature of actual data.

An important aspect of Partial Information Decomposition is the need for selecting a redundancy function from many existing candidates~\cite{Kolchinsky2022}. Our choice of using MMI in the present work was mainly driven by simplicity and low computational cost. However, this redundancy function led to an overestimation of redundancy by considering that the information of the partition with lower mutual information was completely redundant. Thus, MMI sets a lower bound to the synergistic and unique information. For our specific application, we concluded that, among the four $\Phi$-ID atoms that compose the Transfer Entropy measure, the two which are related to redundancy do not substantially contribute to the reconstruction of the network. Therefore, our reconstruction was carried out using the more `conservative' estimation of the relevant atoms. Conservative here means that we are minimizing the estimated information of the relevant atoms. Through this reasoning, we can speculate that redundancy functions which avoid overestimation of redundancy should allow us to obtain even better results. A full exploration of the implications of different redundancy functions to this specific application will be explored in a future work.

With these considerations in mind, in the present study we showed that $\Phi$-ID can be used to infer aspects of the connectivity of biological neuronal networks. First, we assessed the performance of $\Phi$-ID by comparing the structure inferred from different atoms, in isolation or in different combinations, to the weight matrix derived from a dynamic simulation of a biologically-realistic neuronal network. In this process we were able to identify the relevant information atoms for structure inference. Indeed, Transfer Entropy (TE) contains, besides the actually information transfer that we are interested in, other information atoms which, for the aim of inferring networks structure from data, contribute as residuals. $\Phi$-ID allowed us to decompose the TE measures into different atoms to next consider those atoms that were of relevance. By considering the data from numerical simulations (with known ground-truth topology), we showed that the combination of $I_{\partial}^{U^{1}\to U^{2}}$ and $I_{\partial}^{S\to U^{2}}$, i.e. the {\em unique} and {\em synergistic} information transfers, respectively, best captured network structure from spontaneous activity. 

Simulations also provided a criterion to determine the critical threshold below which the false positive rate increases abruptly, and we identified a method to determine this threshold from recorded activity only.
After obtaining the insights from the simulated data, we defined reasonable threshold values for accepting connections as significant, altogether allowing us to apply the $\Phi$-ID framework to analyze actual biological data obtained from spontaneous activity recordings in neuronal cultures. The results showed that, by using $\Phi$-ID, qualitative aspects of the networks could be extracted, better than by using TE directly. Indeed, $\Phi$-ID was able to reveal key topological features in neuronal cultures associated with imprinted anisotropies, in the form of parallel topographical bands, that promote the development of connections along the direction of the bands. 

We note that our use of $\Phi$-ID to infer the topology of living neuronal networks from activity data must be taken with caution. In our case, as in other approaches, the reconstructed effective network can be seen as a proxy of the underlying structure, but the degree of accuracy is very difficult to determine unless complementary tools are taken into account. For instance, Orlandi {\em et al.,} suggested that stimulation is required to further improve predictability~\cite{orlandi2014transfer}. In our case, it would be interesting to extend $\Phi$-ID to take into account direct stimulus response data from neurons and to analyze how intrinsic spatial constraints, which limit the connectivity layout of the network under study, could enhance reconstruction of the network topological core.

$\Phi$-ID offers a higher flexibility when applying information measures such as Transfer Entropy to ascertain specific properties of a system. Selecting the right information atoms helps us to focus on the direct interaction between elements of the system, which is more closely related to the structural topology of the network. Specifically, we showed that the structure inferred by $\Phi$-ID agrees with the known structural aspects of the neuronal culture, even when our measures still capture functional aspects of the network. From the methodological point of view, the analysis also allowed us to select a minimal acceptable threshold, which is determined as the lowest threshold for which a slight decrease in it leads to a rapid increase in wrongly identified links in the reconstructed structure.

With the aim to advance in better and more reliable reconstructions of network connectivity from experimental data, it would be interesting to explore {\em in silico} the impact of excitatory-inhibitory balance and assess how different dynamic regimes derived from the same ground-truth affect reconstruction. This and other challenges provide an exciting opportunity for the application of new analysis tools such as $\Phi$-ID which could help unveiling some of the secrets of complex systems' structure and dynamics.

Finally, the reader should note that the $\Phi$-ID framework is a multivariate extension of Partial Information Decomposition (PID)~\cite{Barrett2015} and can also incorporate the methods of pointwise or local information dynamics (LID)~\cite{Lizier2013} as is done to calculate the integrated information in cellular automata models~\cite{Mediano2022}. In fact, in our application, we are already taking advantage of local information dynamics when, instead of analyzing the whole system with its $N$ elements (neurons or ROIs), we study only the `local' information dynamics of the system by measuring the mutual information atoms of each possible pair $i,j$ of elements. In this sense, the application of $\Phi$-ID to reconstruct the neuronal topology from activity data integrates ideas from both LID and PID, showing the flexibility and generality of the framework. In our particular case, we applied the framework to boolean time series, which simplify the estimation of probability distributions. When working with continuous time series, different strategies can be applied, e.g. the use of entropy estimators such as the ones based on k-neighbors methods ~\cite{Kraskov2004} or linear-gaussian estimators as the ones applied for blood-oxygen-level-dependent (BOLD) signals in~\cite{Luppi2022} to compute mutual information. However, we should bear in mind that there are no estimators free of biases and limitations~\cite{Xiong2017}, specially when dealing with limited continuous variables such as the ones obtained from real complex systems. Nonetheless, these limitations are common to any information theory measure, not only to the framework applied in this work.

In conclusion, our work shows that $\Phi$-ID provides a suitable environment to extract interesting structural topological traits of the connectivity in neuronal cultures from spontaneous activity data. $\Phi$-ID is general and can be applied to other complex systems in which the underlying layout of interactions is unknown, such as gene and protein interactions, species interactions within ecosystems or climate. Conceptually, by extracting the {\em unique} contribution in $\Phi$-ID, one can reveal the backbone of such systems, study vulnerabilities, or make predictions on their evolution.

\begin{acknowledgments}
J.J.T. acknowledges financial support from the Consejería de Transformación Económica, Industria, Conocimiento y Universidades, Junta de Andalucía, Spain and European Regional Development Funds, Ref. P20\_00173. This work is also part of the Project of I+D+i, Ref. PID2020-113681GBI00, financed by MICIN/AEI/10.13039/501100011033 and FEDER, Spain, “A way to make Europe”. J.S. and A.M.H. acknowledge financial support from the European Union’s Horizon 2020 research and innovation program under grant agreement No 964877, project NEU-ChiP; by the Ministerio de Ciencia e Innovación, Spain, Grants
PID2019-108842GB-C21 and PID2022-137713NB-C22; and by the Generalitat de Catalunya, Grant 2021-SGR-00450. G.M. would like to thank the {\em Programa Nacional de Becas de Postgrados en el Exterior ``Don Carlos Antonio López" (BECAL)-Paraguay } for the financial support to his doctoral studies in the Physics and Mathematics Program of the University of Granada. G.M also thanks Yasmin May V. for English writing suggestions.
\end{acknowledgments}

\section*{Data Availability Statement}

The experimental and numerical data used in this study are openly available in Mendeley Data ({\em Neuronal cultures on PDMS topographical patterns: experiments and simulations}, https://doi.org/10.17632/
8yp4xb6d3s.1).

\appendix \label{Sec:Appendix}

\section{Experimental methods} \label{sec:Aexpt}

Neuronal primary cultures were prepared from rat embryonic cortical tissue following the protocol described in Montal\`{a}-Flaquer \emph{et al.}~\cite{montala2022rich}. Briefly, cortical tissue was dissociated by repeated pipetting and neurons seeded on a polydimethylsiloxane (PDMS) circular surface $6$~mm in diameter that contained topographical motif shaped as parallel bands $300$~$\mu$m wide, $70$~$\mu$m high, and separated by $300$~$\mu$m. Neurons were seeded at day {\em 
 in vitro} (DIV) 0 and transduced with the genetically encoded calcium indicator GCaMP6s at DIV 1, and were incubated at 37$^{\circ}$C, 95\% humidity and 5\% CO$_2$ for about 2 weeks with periodic culture medium replacements.  Cultures contained 80\% excitatory neurons and 20\% inhibitory ones in their nature stage. Spontaneous activity was recorded through calcium fluorescence imaging in an inverted microscope in combination with a high speed camera that captured images at 20 ms intervals with a spatial resolution of $5.9$ $\mu$m/pixel. 
 
 Acquired images were analyzed with the software NETCAL~\cite{orlandi2017netcal} to select regions of interest (ROIs), extract their fluorescent signal as a function of time and identify peaks in fluorescence that revealed neuronal spiking events. A total of 900 ROIs were considered in the experiments, which covered in a grid-like manner the area of the culture~\cite{montala2022rich}. ROIs were used instead of single neurons since the large size of the cultures in combination with limitations in camera resolution, made not possible to resolve single cell bodies. An ROI contained about $3-5$ neurons, and 900 ROIs were used to balance spatial details and a reasonable analysis time. The trains of identified spikes, extended to all regions of interest, shaped the raw data from which Transfer Entropy was computed~\cite{stetter2012,montala2022rich}.

The rodents used in these experiments were provided by the animal farm of the University of Barcelona. Their manipulation and tissue dissection were carried out under ethical order B-RP-094/15–7125 of 10th July 2015, which was approved by the Ethics Committee for Animal Experimentation of the University of Barcelona in accordance to the regulations for animal experimentation of the Generalitat de Catalunya (Catalonia, Spain). 

\section{Numerical simulation} \label{sec:Anum}
Numerical simulations followed Orlandi \emph{et al.}\cite{orlandi2013}, in which axon growth and neuronal dynamics were simulated to emulate the spontaneous activity patterns of biological neuronal networks. Simulations in the present study were modified to take into account the influence of topographical bands structure on the direction of axonal growth, replicating the experimental results of Montal\`{a}-Flaquer \emph{et al.}\cite{montala2022rich}.

The connectivity used in the numerical simulations was established as follows.
The $N\approx 2800$ neurons (of which $80\%$ were excitatory) were distributed uniformly on a circular area with radius $r=1.5~\text{mm}$.
Following, axon growth was simulated to determine the projections made by each neuron. 
From the center of a neuron $i$, line-segments were concatenated until the axon reached a certain length $L_i$. The axon lengths $L_i$ were drawn from a Rayleigh distribution with scale parameter $\sigma_L = \sqrt{2/\pi}$, such that the average axon length was $\langle L \rangle = 1~\text{mm}$.
Each consecutive line-segment was placed starting from the end of the previous segment with a random deviation in angle of $\phi=0.1~\text{radians}$.
Once the axon of a neuron $i$ grew into an area of radius $150~\mu\text{m}$ around another neuron $j$, a connection from $i$ to $j$ was formed with a $50\%$ probability. 
The band obstacles influenced the growth of axons by allowing line segments to cross the obstacle borders with a low probability, and with the complementary probability to redirect the line segment to remain in the originating band.

Neuron dynamics were modelled as Izhikevich regular spiking neurons~\cite{izhikevich2003},
\begin{equation}
\begin{aligned}
        \frac{dV_i}{dt} &= 0.04V_i^2 + 5V_i + 140 - U_i + \sum_{j=0}^N w_{ij} P_j + \sigma \eta_i \\
    \frac{dU_i}{dt} &= 0.02(0.2 V_i - U_i),
\end{aligned}
\end{equation}
where $V_i$ is the membrane potential, $U_i$ a recovery variable, and $\eta_i$ a Gaussian white-noise, $\langle \eta_i(t), \eta_i(\tau) \rangle = \delta(t-\tau)$. 
The weighted and directed connectivity matrix $W=\{w_{ij}\}$ resulted from the network-growth algorithm described above.
The neurons were coupled through the variables $P_i$, which evolve following
\begin{equation}
\begin{aligned}
    \frac{dP_i}{dt} &= -\frac{P_i}{\tau_P} + \beta \sum_{k=1}^{N_i} \int_{-\infty}^t  R_i(t') \delta(t'-t_i^k) dt',\\
    \frac{dR_i}{dt} &= \frac{1-R_i}{\tau_R} - \gamma \sum_{k=1}^{N_i} \int_{-\infty}^t R_i(t') \delta(t'-t_i^k) dt',
\end{aligned}
\end{equation}
where the index $k=1,\ldots,N_i$ enumerates all the spikes emitted by neuron $i$, $t_i^k$ is the time of the $k$-th spike of neuron $i$, $\beta, \gamma < 1$ are two constants and $\tau_P < \tau_R$ the time-constants of the synaptic dynamics.

\section{Calculation of graph measures}
The graph theoretic measures used in Figs. \ref{fig:Sim_recons} and \ref{fig:Exp_recons} were computed using the Brain Connectivity Toolbox~\cite{rubinov2010}. For {\em in vitro} data, these measures were computed on the effective connectivity matrices $A=\{a_{ij}\}$, while for {\em in silico} data, these measures were computed on either $W=\{w_{ij}\}$ (ground truth topology) or $A=\{a_{ij}\}$. For clarity, the matrix $A$ was used in the definitions below.

\subsection{In-degree}
The distribution of in-degrees $p(k_{in})$ in effective networks captured the probability to observe a node in the network (neurons or ROIs) with $k_{in}$ incoming links. For a given node $i$, its number of incoming links was calculated by summing up the number of existing directed links toward the node, as 
\begin{equation}
    k_{in}^{i} = \sum_j a_{ij},
\end{equation}
where $a_{ij}$ are the elements of the effective connectivity matrix $A$, i.e. the set of directed links in the network, with $j \rightarrow i$ indicating that there exist an effective link from neuron $j$ to $i$.

\subsection{Global efficiency $G_{E}$}
The global efficiency $G_{E}$ quantified the easiness for network-wide integration, i.e., the easiness for a node in the network to topologically reach any other node. $G_{E}$ was defined as the average of the inverse of the pair-wise topological network distances~\cite{rubinov2010}, as
\begin{equation}
    G_{E} = \frac{1}{N(N-1)} \sum_{(i,j) \in A} \frac{1}{d(i,j)},
\end{equation}
where $d(i,j)$ is the shortest geodesic path connecting $i$ to $j$ in the matrix $A$, with $(i,j) \notin A \implies d(i,j) = \infty$, and $N$ denotes the number of neurons. A value $G_E=0$ reflected a network in which all nodes were disconnected, while $G_E=1$ reflected a complete graph, i.e., a network in which any given node connected to all its possible $N-1$ neighbors. 

$G_E$  was used as a network metric to characterize the topological organization of the reconstructed effective networks in a quantitative manner and for different thresholds, since its value intuitively accounts for the density of links in a network.

\subsection{Modularity Q}
The modularity Q measured the tendency of neurons to connect with other neurons within a group or community rather than with other neurons in the rest of the network~\cite{rubinov2010}. It was calculated as
\begin{equation}
    Q = \frac{1}{2m}\sum_{i,j \in M} \left( a_{ij} - \frac{k_i k_j}{2m}\right) \delta(c_i,c_j),
\end{equation}
where $M=\{1,...,N\}$ is the set of neurons in the network, $a_{ij}$ are the elements of the effective connectivity matrix $A$, $m=(1/2)|a|$, $k_i=\sum_{j} a_{ij} + \sum_{j} a_{ji}$ is the total degree ($k^i=k^i_{in}+k^i_{out}$) of neuron $i$, $c_i$ denotes the community that neuron $i$ belongs to, and $\delta(\cdot,\cdot)$ is the Kronecker delta such that
\begin{equation*}
    \delta(c_i,c_j) = \begin{cases} 
            1, & c_i = c_j \\
            0, & \text{otherwise}.
        \end{cases}
\end{equation*}

Communities were detected using the Louvain algorithm~\cite{blondel2008fast}. $Q$ was used throughout the analysis as a network measure to quantify the tendency of nodes (neurons or ROIs) to organize into communities, with $Q=0$ indicating that the entire network was the only community and $Q = 1$ indicating that all nodes were isolated and formed by themselves a community. The interest of using $Q$ and the Louvain algorithm was also based on the fact that they reflected well the organizational features of the {\em in silico} networks, in which neurons connected along preferred directions with clear communities, and therefore were considered adequate to later compare the characteristics of reconstructed {\em in silico} and {\em in vitro} data. However, the used modularity measure could not work adequately with other data, and other definitions could therefore be considered~\cite{diez2015novel,MacMahon2015,Newman2016}

\nocite{*}

\bibliography{aipsamp}
\end{document}